# Edge-Dominated Twist Mechanics at van der Waals Interfaces

Yifan Shao[1#], Zhaoheng Zhang[2#], Hao Li[1], Oded Hod[3*], Michael Urbakh[3*], Quanshui Zheng[1,4], Xiang Gao[2*], Deli Peng[1,4*]

[1]*Center of Double Helix, Institute of Materials Research, Tsinghua Shenzhen International Graduate School, Tsinghua University, Shenzhen, China*

[2]*Department of Modern Mechanics, School of Engineering Science, University of Science and Technology of China, Hefei 230027, Anhui, China*

[3]*Department of Physical Chemistry, School of Chemistry, The Raymond and Beverly Sackler Faculty of Exact Sciences and The Sackler Center for Computational Molecular and Materials Science, Tel Aviv University, Tel Aviv 6997801, Israel*

[4]*Institute of Superlubricity Technology, Research Institute of Tsinghua University in Shenzhen, Shenzhen, China*

[#]These authors contributed equally to this work.

[*]Contact author: odedhod@tauex.tau.ac.il; urbakh@tauex.tau.ac.il; xianggao@ustc.edu.cn; pengdeli@sz.tsinghua.edu.cn

## Abstract

Despite the pivotal role of twist in modulating physical properties at van der Waals (vdW) interfaces, the mechanics governing torsional response remain poorly understood. Here, we probe twist mechanics at homo- and heterogeneous vdW interfaces, together with their sliding behaviors within a unified experimental framework. For both systems, the peak torque scales nearly linearly with contact area, in contrast to predictions from linear elastic and rigid models. Remarkably, while the sliding friction of the two interfaces diverges by over three orders of magnitude owing to different scaling laws, the corresponding torque follows the same linear scaling and differs by only about twenty-fold. Large-scale atomistic simulations reveal an edge-dominated yielding mechanism for torsional motion, wherein elastic reconstruction shifts the effective load-bearing region toward the edges, eliminating torque from the contact interior. This mechanism contrasts with the bulk-mediated stress transmission governing translational sliding, a distinction rooted in the different loading geometries inherent to the two motion modes, where torsional loading necessitates perimeter actuation, whereas sliding enables center-driven loading. This symmetry-imposed divergence demonstrates that translational and torsional properties cannot be predicted from one another at vdW interfaces, providing critical insights for the design of dynamically reconfigurable micro- and nanoelectromechanical devices.

## Introduction

The controlled twist of stacked two-dimensional (2D) crystals has emerged as a powerful strategy for tuning the physical properties of vdW interfaces[1–3]. By introducing a relative misorientation between atomic lattices, twist generates moiré superstructures that host a wide range of emergent electronic, optical, quantum, mechanical, and tribological phenomena[4–13]. Despite this central role, twist is predominantly treated as a fixed structural parameter once device fabrication is complete, while its dynamic mechanical responses remain largely unexplored. Elucidating the torsional mechanics at vdW interfaces is essential not only for controlling the stability of moiré structures[14,15], but also for advancing reconfigurable twistronic systems, where interfacial states must be dynamically and precisely tuned *in situ*[16–20].

Recent studies of rotational dynamics and torsional behaviors at vdW interfaces have revealed spontaneous alignment[21–23], twistable device architectures[24,25], measurable torsional responses[16,26,27], as well as stick-slip transition and emerging strain chirality[28,29]. However, the physical mechanisms governing torsional motion remain largely unresolved, particularly how interfacial strain and stress evolve during yielding and how torsional behaviors (e.g., torque) relate to translational quantities (e.g., shear stress and friction).

In classical continuum mechanics, under the assumption of homogeneous linear elasticity[30], torque is proportional to the shear modulus $G$, the torsional constant $J$, and the twist angle $\theta$, i.e., $T \propto GJ\theta$. At vdW interfaces, the torsional constant $J$ equals the second moment of area due to an invariant cross-section and negligible warping. This description consequently predicts a superlinear scaling of torque with contact area, $T \propto A^2$ for given $G$ and $\theta$, since $J \propto A^2$. However, this classical picture neglects the complex elastic reconstruction induced by registry-dependent anisotropic interlayer interactions at vdW interfaces[31], which gives rise to highly nonuniform interfacial stress distributions[14], calling into question the applicability of linear elastic theory.

In this work, we investigate twist mechanics at vdW interfaces using a unified experimental platform that enables direct measurements of both torsional and translational forces in homo- and heterostructures. We find that the peak torque scales approximately linearly with contact area in both systems, in contrast to classical model predictions. Large-scale atomistic simulations reveal that this discrepancy originates from an edge-dominated yielding mechanism, in which atomic reconstruction confines the loading-bearing region to the edges, thus suppressing contributions from the contact interior. Comparison with translational sliding further demonstrates that these two motion modes activate distinct evolution pathways of stress transmission, traced to their inherently different loading geometries. Our findings establish an edge-yielding paradigm that governs twist mechanics at vdW interfaces, providing a foundation for the rational design of reconfigurable twistronic devices.

**Main Text**

We investigate the torsional and translational mechanical responses of both incommensurate and commensurate vdW interfaces using single-crystal *h*-BN/graphite (Gr) heterostructure and graphite/graphite (Gr/Gr) homostructure mesas. Cylindrical mesas with integrated cantilever arms were fabricated and manipulated *in situ* using an atomic force microscope (AFM). To directly compare torsional and translational mechanics within a unified framework, our experimental setup enables two different manipulation modes: applying a tangential force at the lever arm to twist the top mesa while preserving a fixed contact area (Figure 1a, see Supplementary Fig. 1 for fabrication and manipulation details), and applying a lateral force on the metal cap to slide the top mesa forward and backward (Figure 1b, and Supplementary Video 1). During torsional manipulation, the top mesa undergoes stable concentric rotation without apparent lateral displacement, as confirmed by consecutive AFM topography images (Figure 1c) and real-time optical imaging (Supplementary Video 2). This configuration isolates nearly pure interfacial torsion, suppressing translational motion and thereby enabling a direct investigation of torsional mechanics.

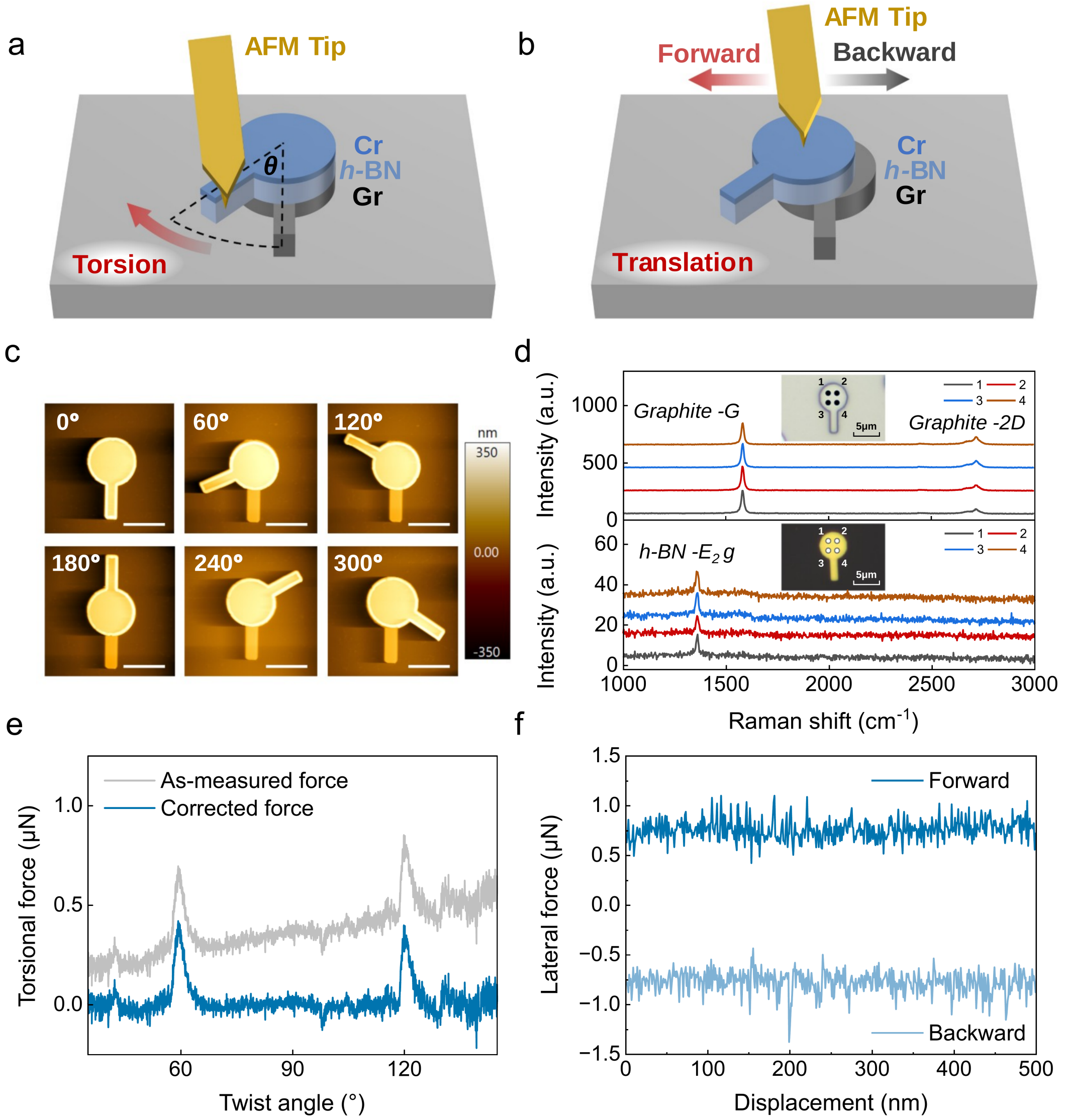

**Figure 1. Torsional and lateral force measurements of vdW mesas.** (a)-(b) Schematics of the experimental setup of an *h*-BN/Gr mesa for *in situ* manipulation using an AFM tip: (a) applying a tangential force to twist the top *h*-BN mesa relative to the underlying graphite mesa, and (b) applying a lateral force to slide the top *h*-BN mesa back and forth. (c) AFM topography images of a same stacked mesa acquired at different twist angles during rotation. Scale bars, 5 μm. (d) Raman spectra of the exposed bottom graphite mesa and the backside of the transferred *h*-BN mesa. (e) Torsional forces as a function of twist angle for a 2-μm-radius mesa at an angular velocity of 18 °/s, showing both the as-measured and the baseline-corrected curves. (f) Lateral force loops acquired in forward and backward sliding with the same mesa in panel (e) at a sliding velocity of 0.15 μm/s.

To verify that rotation motion occurs at the heterointerface of $h$-BN/Gr, we performed Raman spectroscopy on an $h$-BN/Gr heterostructure, examining both the exposed bottom surface and the backside of the transferred top mesa. As shown in Figure 1d, the bottom surface exhibits graphite $G$ and $2D$ peaks, while the top mesa shows the characteristic $h$-BN $E_{2g}$ mode, with no detectable defect-related features. AFM topography further confirms atomically smooth surfaces (Supplementary Fig. 2). These results indicate that rotation indeed occurs at the clean and atomically flat heterointerface of $h$-BN/Gr.

Figure 1e presents representative torsional force profiles as a function of twist angle for an $h$-BN/Gr mesa (see Supplementary Fig. 3 for torsional force extraction). The as-measured torsional force typically contains low-frequency baseline drift; once this background is subtracted, the corrected torsional force clearly reveals pronounced torque peaks with 60° periodicity, reflecting the shared hexagonal lattice symmetry of $h$-BN and graphite. The peak torque shows no discernible dependence on angular velocity within the investigated range (Supplementary Fig. 4), indicating a negligible thermal activation effect on the crossing of the high rotational energy barriers. Moreover, we performed corresponding sliding measurements with the same mesa system. Figure 1f displays the lateral force traces acquired by sliding the mesa under a 60° aligned configuration. These results validate the mesa system as a comprehensive platform to explore both torsional and translational interfacial mechanics.

We next quantitatively compare the torsional responses of the homo- and heterointerfaces. Figure 2a shows the torque-angle curves for the Gr/Gr and $h$-BN/Gr systems with the same radius of 2 μm. The peak torque of the Gr/Gr homointerface exceeds that of the $h$-BN/Gr heterointerface by one order of magnitude. To characterize the size dependence, we extract peak torque as a function of contact area. In both systems, peak torque scales linearly with contact area, $T^{\max} \propto A$, yielding well-defined areal torsional strengths of 3.32 and 0.17 $N \cdot m^{-1}$ for the homo- and heterointerfaces, respectively (Figure 2b). Notably, this linear scaling qualitatively contradicts the superlinear scaling of $T \propto A^2$ predicted by classical continuum mechanics under the linear elastic assumption.

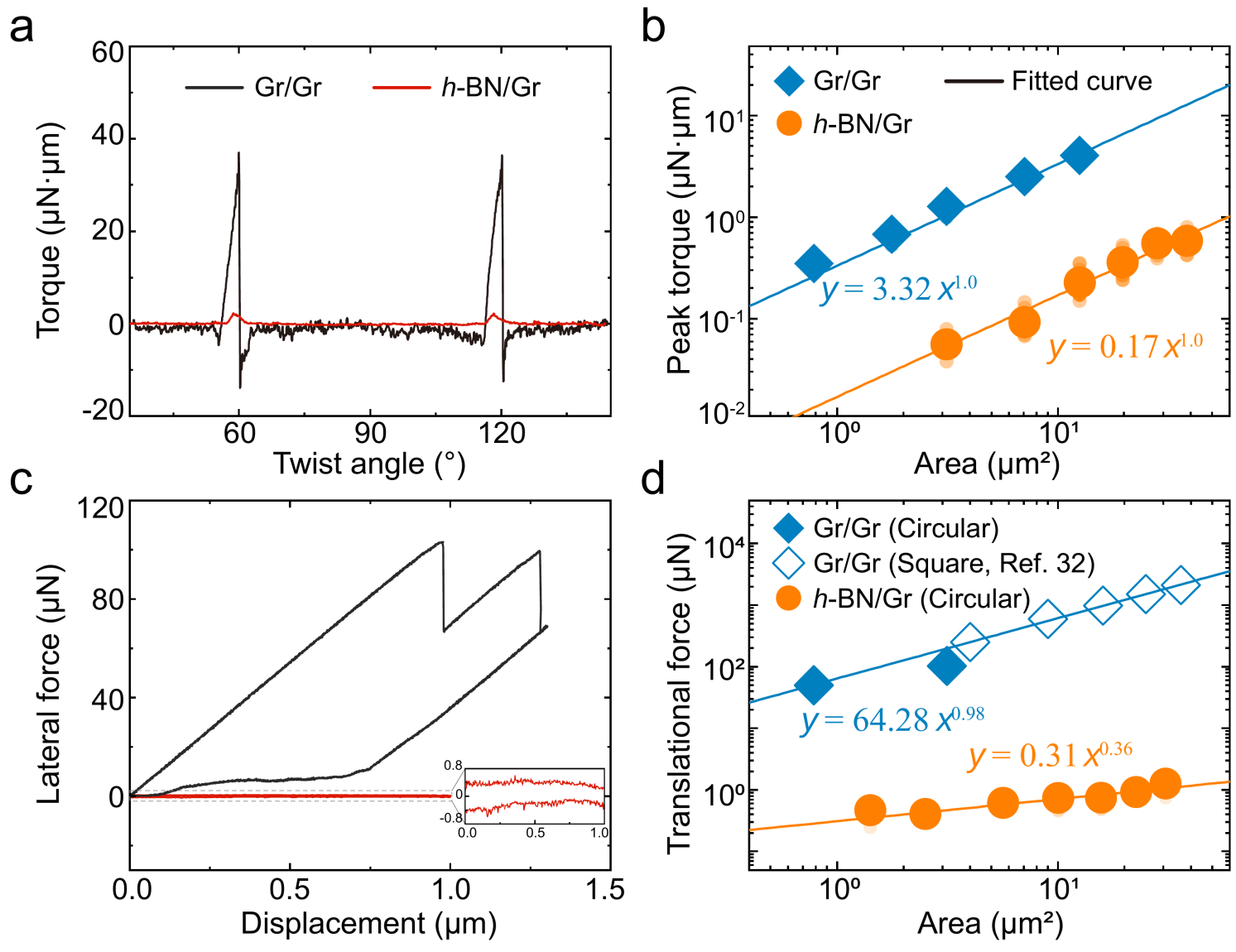

**Figure 2. Comparison of torsional and translational responses at vdW interfaces.** (a) Torque profiles as a function of twist angle for Gr/Gr (black line) and *h*-BN/Gr (red line) interfaces of the same size ($R \approx 2$ μm). (b) Scaling of peak torque with contact area for Gr/Gr and *h*-BN/Gr interfaces. (c) Lateral force traces as a function of displacement for Gr/Gr and *h*-BN/Gr interfaces of the same size ($R \approx 1$ μm). (d) Scaling of translational force with contact area for Gr/Gr and *h*-BN/Gr interfaces. The translational forces for square Gr/Gr mesas are adapted from previous work[32].

To assess the relationship between torsional and translational responses, we measured the lateral force traces for same-sized Gr/Gr and *h*-BN/Gr interfaces (Figure 2c, and Supplementary Note 3). The translational force was defined as the maximum lateral force during the initial loading stage for Gr/Gr, and as half of the maximum difference between the forward and backward traces for *h*-BN/Gr. The commensurate Gr/Gr interface exhibits characteristic stick-slip behavior consistent with previous reports[32], yielding a translational force that is more than two orders of magnitude higher than that of the aligned incommensurate *h*-BN/Gr heterointerface for $R \approx 1$ μm. More importantly, the two types of interfaces exhibit distinct area-scaling behaviors, as shown in Figure 2d. The commensurate Gr/Gr homointerface displays a nearly linear scaling, whereas the incommensurate $h$-BN/Gr heterointerface shows a sublinear scaling with an exponent of ~0.36. This scaling exponent is close to the theoretical prediction of 0.25 for circular flakes, which is attributed to edge-dominated incomplete moiré tiles therein[33].

Taken together, twist mechanics at vdW interface present distinct behaviors compared to slide mechanics. While the translational force of the homo- and heterointerfaces differs by more than three orders of magnitude (for $R > 2$ μm) with different scaling laws, the corresponding peak torque of the two systems follows the same linear scaling and differs by only a factor of ~20. These results suggest that translation and torsion are governed by different mechanisms at vdW interfaces.

To provide a theoretical benchmark, we first adopt the classical linear-elastic framework and assume that rotational slip occurs when the local shear stress at the contact rim reaches a critical value of the dynamic shear stress $\tau_{\mathrm{s}}$. Under this assumption, a simple estimate for the peak torque of a circular vdW contact can be derived as

$$T_{\mathrm{cir,lin}}^{\mathrm{max}}(R) = \int_0^R \int_0^{2\pi} \frac{r}{R} \tau_{\mathrm{s}} r^2 d\theta dr = \frac{1}{2} \pi \tau_{\mathrm{s}} R^3. \quad (1)$$

Substituting the previously measured interfacial shear stress of a commensurate graphite interface, $\tau_{\mathrm{s}} \approx 62$ MPa[32], Eq. (1) gives a peak torque of 779 μN·μm for $R \approx 2$ μm, which, however, is nearly 20 times larger than the experimentally measured value. Moreover, the predicted area scaling $T_{\mathrm{cir,lin}}^{\mathrm{max}} \propto A^{1.5}$ contrasts with the experimentally observed linear scaling.

A more rigorous treatment, based on a continuum description of the moiré potential energy landscape for a rigid interface, yields the following expression for the torque of a circular flake[29,33]:

$$T_{\mathrm{cir,rig}}(\boldsymbol{x}_0, \theta, R) = -\frac{4}{9} p \pi U R^2 \frac{\lambda(\theta)^2 \sin\theta}{a_{\mathrm{t}} a_{\mathrm{b}}} J_2\left(\frac{4\pi R}{\sqrt{3}\lambda(\theta)}\right) f(\boldsymbol{x}_0), \quad (2)$$

where $\boldsymbol{x}_0$ is the center-of-mass of the flake, $U$ is the amplitude of the potential energy landscape corrugation per unit area, $\lambda(\theta)$ is the moiré period, $J_2$ is the second-order Bessel function of the first kind, $a_{\mathrm{b}}$ and $a_{\mathrm{t}}$ are the lattice constants for the bottom and top layers, respectively, giving the lattice misfit parameter $\delta = a_{\mathrm{b}}/a_{\mathrm{t}} - 1$, and $f(\boldsymbol{x}_0)$ is a position-dependent term. The prefactor $p$ takes the value $\pm 1$ for homointerfaces and heterointerfaces, respectively. By applying small-angle approximation ($\theta \ll 1°$) for $\delta = 0$ and asymptotic expansion ($R \gg 1$ nm) for $\delta \neq 0$, we obtain the peak torque $T_{\mathrm{cir,rig}}^{\mathrm{max}}$ and establish the explicit scaling laws for low-energy stacking (AB stacking at the center):

$$T_{\mathrm{cir,rig}}^{\mathrm{max}}(A) \approx \begin{cases} \frac{0.277\pi^{0.5}U}{a_{\mathrm{t}}} A^{1.5}, \delta = 0 \\ \frac{0.535 a_{\mathrm{b}} U}{\pi^{0.75}\delta^{1.5} a_{\mathrm{t}}^{0.5}} A^{0.75}, \delta \neq 0 \end{cases}. \tag{3}$$

See Supplementary Note 5 for further details of model derivation.

The rigid model predicts a torque scaling of $A^{1.5}$ for Gr/Gr homostructure, similar to the linear elastic model, whereas it predicts a weaker $A^{0.75}$ dependence for *h*-BN/Gr heterostructure, close to the experimentally observed linear scaling. Overall, disagreement remains between these theoretical predictions and the experimental observations, particularly for the Gr/Gr homostructures, suggesting that neither the linear elastic nor rigid descriptions can adequately capture the twist mechanics at vdW interfaces.

To examine the intricate effect of elasticity at atomic level, we performed quasistatic molecular dynamics (MD) simulations with a tri-layer flexible model system as illustrated in Figure 3a (See Methods for further simulation details). To highlight the effect of elasticity, additional atomistic calculations with a rigid flake on a rigid substrate were also performed. Figure 3b presents the torque–angle profiles for a 20-nm-radius flake obtained from the rigid model (black line, Eq. 2), rigid atomistic calculations (open circles), and flexible quasistatic simulations (red line). The rigid atomistic calculations agree well with the rigid model predictions for both homo- and heterostructures. By contrast, elastic relaxation dramatically alters the torsional response, giving rise to stick-slip behavior and markedly lower peak torque than the rigid scenarios, particularly for the homointerface.

Remarkably, Figure 3c demonstrates that the peak torque obtained from flexible

MD simulations exhibits linear scaling with contact area for both Gr/Gr and *h*-BN/Gr interfaces, consistent with the experimental observations. The extracted areal torsional strengths are 2.01 $N \cdot m^{-1}$ for the Gr/Gr homointerface and 0.63 $N \cdot m^{-1}$ for the *h*-BN/Gr heterointerface. Although the simulated strength ratio between the two systems is several times lower than that in larger-scale experiments, the same qualitative trend is captured. Moreover, these results contrast with the rigid model predictions and calculations (inset of Figure 3c), highlighting the essential role of elastic reconstruction in determining torsional response.

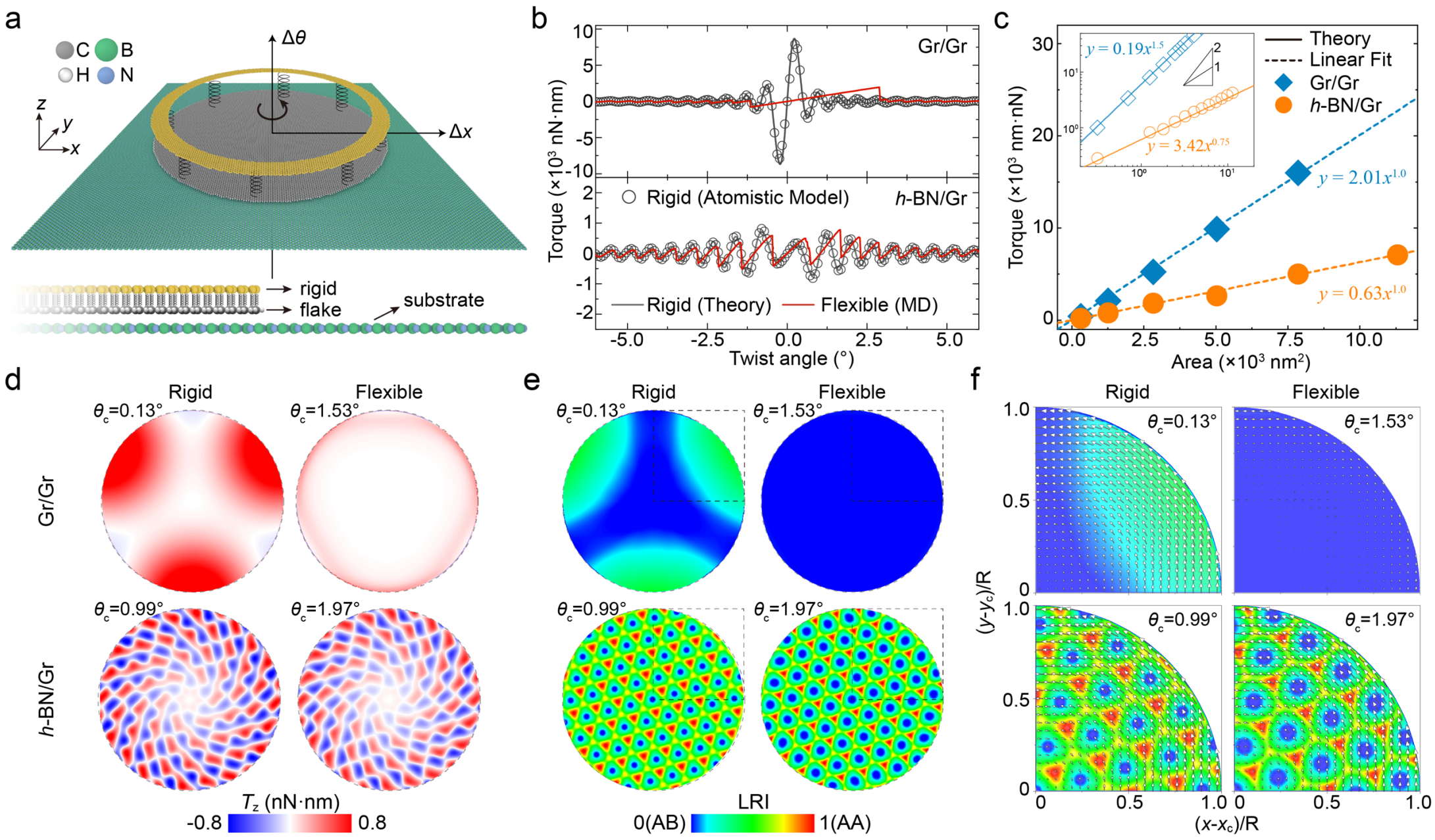

**Figure 3. Mechanism of twist-induced torque at vdW interfaces.** (a) Schematic of the flexible MD simulation system. Top panel: perspective view of a circular flake driven by a rigid ring with effective interlayer springs. Bottom panel: cross-sectional view of the atomic configuration. (b) Simulated torque profiles as a function of twist angle for Gr/Gr and *h*-BN/Gr interfaces with a 20-nm-radius flake, obtained from the rigid model (black line, Eq. 2), rigid atomistic calculations (empty circles), and flexible quasistatic MD simulations (red lines). (c) Scaling of peak torque with contact area (solid symbols: quasistatic MD results; dashed lines: linear fitting) for Gr/Gr and *h*-BN/Gr interfaces. The inset displays the results of rigid model (solid lines, Eq. 3) and rigid atomistic calculations (void symbols). (d)-(e) Spatial distributions of (d) the local

torque per atom ($T_z$) and (e) the LRI for the rigid and flexible interfaces at the peak torque state with the critical twist angle ($\theta_c$). For clarity, the torque value of each atom is averaged over its nearest neighbors. (f) The atomic displacement fields corresponding to the top right quarters of the LRI maps in panel (e) with respect to their equilibrium states. Results in panels (d)-(f) are taken from simulations with a 40-nm-radius flake.

To unveil the origin of the reduction in peak torque and its scaling in the flexible simulations, we analyze the spatial distribution of local torque per atom $T_z$ acting on the flake by the substrate at the peak torque state (Figure 3d). For Gr/Gr homostructure, the rigid model exhibits a widespread $T_z$ distribution, whereas the flexible model generates torque mainly along the periphery with negligible contribution from the interior region. The underlying stacking evolution is visualized using the local registry index (LRI)[34–36] in Figure 3e. The high-torque sites in the rigid model correspond to the saddle-point stacking domains, whereas the flexible Gr/Gr interface is overwhelmingly occupied by low-energy AB stacking, leaving only a minor degree of lattice misfit within the rim region, which remains visually indistinguishable due to low LRI values (see Supplementary Fig. 7).

The atomic displacement fields in Figure 3f further reveal that at small twist angles, the flexible Gr/Gr system undergoes pronounced atomic reconstruction, wherein the interior of the graphene flake remains locked in the AB stacking configuration, while interfacial displacement mainly occurs at the edge region. This spontaneously eliminates the high-energy saddle-point stacking regions, leading to a dramatic reduction in the peak torque compared to the rigid model predictions. As the torsional loading proceeds, saddle-point stacking nucleates at the rim and propagates inward, abruptly relaxing the commensurate interface upon slip (Supplementary Video 3).

For $h$-BN/Gr heterostructure, the aligned configuration remains incommensurate owing to the intrinsic lattice mismatch (~1.8%) of the two materials, manifesting dense periodic moiré patterns in both the rigid and flexible models. Within the contact interior, opposite local torque contributions within complete moiré tiles largely cancel one another, such that the net torque is dominated by the incomplete moiré tiles at the edges for both cases. Therefore, despite some differences in yielding twist angle and moiré morphology, the rigid and flexible systems present generally similar distributions of torque, stacking, and atomic displacements (Figure 3d-e).

The presented results support an edge-yielding mechanism that the torsional responses of both homo- and heterostructures are governed not by the contact interior but by an effective load-bearing region confined to the edges. Guided by this, we therefore propose a modified rigid model, in which torsional stress is transmitted only within a characteristic distance $d_0$ from the contact edge before slip (Supplementary Note 6). The resulting torque scaling can be expressed as

$$T_{\mathrm{cir,rig2}}^{\max}(A) \approx T_{\mathrm{cir,rig}}^{\max}(R) - T_{\mathrm{cir,rig}}^{\max}(R-d_0) \propto \begin{cases} A & ,\delta = 0 \\ A^{0.75} & ,\delta \neq 0 \end{cases}. \quad (4)$$

Fitting Eq. (4) to the flexible quasistatic MD simulation results gives $d_0 \approx 1.6$ nm for the homostructure and $d_0 \approx 10.0$ nm for the heterostructure (Supplementary Fig. 6). The former approaches the ring width, while the latter is close to the maximum moiré period of 13.9 nm for $h$-BN/Gr, suggesting that the scaling prefactor for Gr/Gr homostructures is sensitive to the loading range whereas the $h$-BN heterostructures are dominated by the incomplete moiré tiles at the edges. Notably, the torque scaling of $T_{\mathrm{cir,rig}}^{\max} \propto A^{0.75}$ predicted by the modified rigid model and the linear scaling are too close to be definitely distinguished in current experiments (see Supplementary Fig. 6).

To rationalize the pronounced scaling contrast between the translational force and peak torque shown in Figure 2, we performed quasistatic sliding MD simulations with Gr/Gr homo- and $h$-BN/Gr heterostructures, both in the aligned configurations. Considering that the shear loading applied by pushing the top center of the metal cap of a mesa in sliding experiments generates more uniform stress across the metal/Gr and metal/$h$-BN interfaces[32], an additional type of simulation system with a full driving disk is also employed to investigate the effect of the distinct loading geometries inherent to torsion and sliding.

The simulated lateral force traces for a Gr flake with $R \approx 30$ nm, presented in Figure 4a, demonstrate one order of magnitude difference in the translational force between the homo- and heterointerfaces, as well as a significant reduction for the homointerface from the full disk-driven scheme to the ring-driven scheme. Moreover, the translational forces extracted exhibit markedly different area-scaling laws, showing $F_{\mathrm{cir}}^{\max} \propto A^{0.535}$ for the Gr/Gr and $F_{\mathrm{cir}}^{\max} \propto A^{0.25}$ for $h$-BN/Gr under the ring-driven scheme, and $F_{\mathrm{cir}}^{\max} \propto A^{1.0}$ for the Gr/Gr and $F_{\mathrm{cir}}^{\max} \propto A^{0.25}$ for $h$-BN/Gr under the full disk-driven scheme (Figure 4b). For both driving schemes, the contrast in the translational force between the two systems enlarges with increasing contact area due to different scaling laws, consistent with the experimental observations. Yet, the scaling laws under full disk-driven show overall better agreement with experiments, supporting that pushing the top center of the metal cap introduces more uniform shear loading.

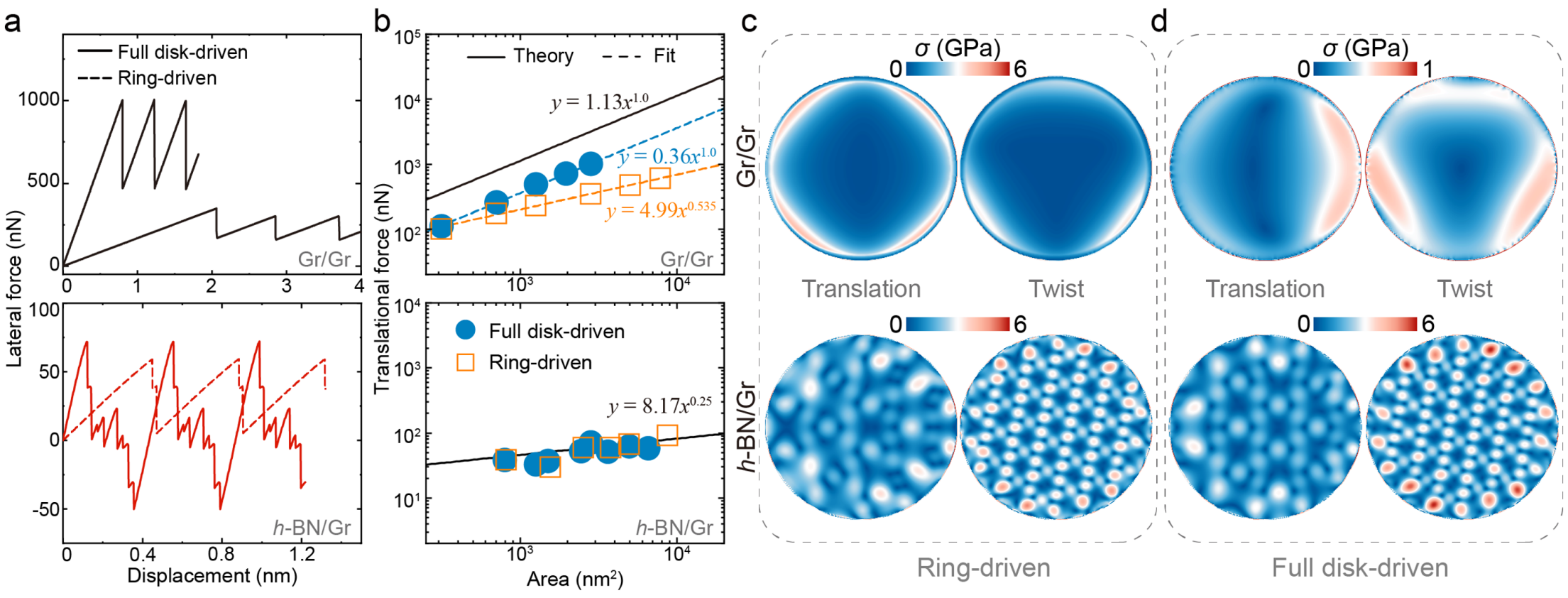

**Figure 4. Simulated translational behaviors and stress comparison with torsion.** (a) Lateral force traces as a function of sliding displacement for flexible Gr/Gr homostructure (upper panel) and $h$-BN/Gr heterostructure (lower panel) with $R = 30$ nm under the ring-driven and full disk-driven schemes. (b) Scaling of translational force with contact area for the Gr/Gr homostructure (upper panel) and $h$-BN/Gr heterostructure (lower panel). Solid lines correspond to the rigid model predictions (Eqs. S14 and S16) and the dashed lines denote power-law fitting. (c)-(d) Spatial distributions of in-plane deviatoric stress ($\sigma$) for Gr/Gr and $h$-BN/Gr interfaces at the translational force (left columns) and peak torque (right columns) states under (c) the ring-driven scheme and (d) under the full disk-driven scheme.

The different scaling behaviors of the translational force for Gr/Gr homointerfaces in Figure 4b lie in the different stress evolution pathways under the two driving schemes. Under ring driving and prior to slip, the in-plane deviatoric stress remains highly concentrated at the edges for both translational and torsional motion modes (Figure 4c). Under full disk driving, the stress field extends from the edges toward the contact interior for both motion modes (Figure 4d). In comparison, the $h$-BN/Gr heterointerfaces show stress distributions that closely follow the moiré superstructure regardless of driving scheme, exhibiting only a modest stress enhancement near the edges. Thus, the $h$-BN/Gr heterointerfaces presents similar scaling laws under both driving schemes.

Overall, these simulations reveal a symmetry-imposed divergence between torsional and translational mechanics. Rotational motion requires perimeter actuation to generate a moment, producing edge-localized stress (ring-driven scheme); translational motion, by contrast, is naturally driven through the contact center, enabling bulk-mediated stress propagation (full disk-driven scheme). This difference in loading geometry, inherent to the symmetries of the two motion modes, explains the contrasting scaling behaviors observed in experiments that edge-dominated stress accumulation in torsion gives rise to linear scaling ( $T^{\max} \propto A$ ), whereas bulk-mediated stress propagation in sliding produces area-dependent scaling ($F_{\mathrm{cir}}^{\max} \propto A$ for commensurate, $F_{\mathrm{cir}}^{\max} \propto A^{0.25}$ for incommensurate interfaces). The distinct stress transmission pathways highlight that the mechanical response of a vdW interface depends not only on interfacial properties but also on the symmetry of the imposed deformation mode.

**Conclusions**

In summary, our results establish an edge-dominated mechanism for twist mechanics at van der Waals interfaces. Rather than arising from linear elastic deformation across the entire contact, the torsional response is governed by an edge-localized load-bearing region that emerges from elastic atomic reconstruction. This picture, together with a modified rigid model, quantitatively explains the observed linear scaling of torque with contact area and rationalizes the essentially different torsional responses of homo- and heterointerfaces. A direct comparison of rotational and translational responses reveals that these two modes of interfacial motion are governed by distinct pathways of interfacial stress transmission that are rooted in their inherently distinct loading geometries, and therefore translational and torsional properties cannot be inferred from one another. This coupling between motion symmetry, loading geometry, and stress transmission pathway provides a general principle for understanding interfacial mechanics in layered materials, and offer design guidance for dynamically reconfigurable micro- and nanoelectromechanical devices.

## Methods

### Fabrication and characterization of stacked vdW mesas

Circular $h$-BN/Gr and Gr/Gr mesas with integrated lever arms were fabricated for *in situ* torsional and translational manipulation. A 50-nm-thick Cr film was deposited by magnetron sputtering as a hard mask. Circular mesa and lever-arm patterns were defined in PMMA by electron-beam lithography. The exposed Cr mask was patterned by ion-beam etching, after which $h$-BN was etched using an $SF_6$-dominant dry-etch process, and thereafter graphite was etched by oxygen plasma. Mesas with radii from 0.5 to 3.5 μm were fabricated and subsequently characterized by SEM and AFM.

### Post-manipulation interfacial characterization

After torsional manipulation, selected top $h$-BN mesas were laterally displaced and lifted using a PDMS stamp to expose the buried interface. AFM topography was acquired on both the separated bottom $h$-BN surface and the exposed top surface of the underlying graphite mesa. Raman point spectra from these surfaces confirmed their material identities via the graphite $G$ and $2D$ bands and the $h$-BN $E_2g$ mode. Raman measurements were performed using a WITec Raman system equipped with a UHTS 300S VIS spectrograph, a 532-nm excitation laser, a 600 grooves $mm^{-1}$ grating, and a 100× objective. Point spectra were acquired with 10 accumulations of 1 s each at a laser power of 2.1 mW.

### Extraction of interfacial torque

During torsional manipulation, the recorded AFM lateral force, $F_{\mathrm{LF}}(\theta)$, corresponds to the projection of the tangential driving force $F_{\tan}(\theta)$ applied to the lever arm onto the lateral-force detection direction,

$$F_{\mathrm{LF}}(\theta) = F_{\tan}(\theta)\cos\theta.$$

The apparent interfacial torque is $\tilde{T}(\theta) = r_{\mathrm{eff}} F_{\tan}(\theta)$, with $r_{\mathrm{eff}}$ the prescribed radius of the circular trajectory followed by the AFM tip. A slowly varying baseline $B(\theta)$ was estimated from $\tilde{T}(\theta)$ by asymmetric least-squares (AsLS) smoothing ($p = 0.001$, $\lambda = 10^9$, 10 iterations) and subtracted to obtain the baseline-corrected torque:

$$T(\theta) = \tilde{T}(\theta) - B(\theta).$$

For each retained peak, the apex was identified and the baseline-corrected trace in a narrow angular window near the apex was fitted with a piecewise-linear function to suppress high-frequency noise. The peak height $\Delta T_{\text{peak}}$ was defined as the difference between the fitted torque at the apex, $T_{\text{fit,max}}$, and the mean baseline-corrected torque in the immediately preceding reference window $W_{\text{pre}}$:

$$\Delta T_{\text{peak}} = T_{\text{fit,max}} - \langle T(\theta) \rangle_{\theta \in W_{\text{pre}}}.$$

Angular intervals, where the projection correction substantially amplified instrumental noise, were excluded from peak-height analysis.

**Measurement of translational force**

Two calibrated platforms were used to accommodate the large difference in the characteristic translational forces between $h$-BN/Gr and Gr/Gr interfaces. Translational forces at $h$-BN/Gr interfaces were measured using the calibrated lateral-force channel of an AFM (NTEGRA Prima, NT-MDT), whereas the substantially larger forces at Gr/Gr interfaces were measured with a calibrated micro-force sensor (FT-S2000, FemtoTools). Both platforms were operated within their calibrated force ranges.

**Molecular dynamics simulations**

The flexible MD system consisted of a rigid driving ring with a width of 2 nm, a flexible finite-sized circular graphene flake with hydrogen-passivated edges, and a flexible Gr or $h$-BN substrate layer (Figure 3a). The rigid driving ring was duplicated from and coupled to the flake carbon atoms via in-plane harmonic springs[29], mimicking the edge actuation by the mesa arm in experiments. For the full disk-driven scheme, the rigid ring was replaced by a rigid flake fully duplicated from the underlying graphene flake. Periodic boundary conditions were applied in the $x$ and $y$ directions. Intralayer interactions were described using the REBO[37] and Tersoff[38] potentials for graphene and $h$-BN, respectively, while interlayer interactions are accounted for by the registry-dependent ILP[39–42].

The in-plane stiffness of the harmonic springs connecting each driving atom to its corresponding carbon atom was set to $k_{xy} = k_{\text{int}}/n$, with $k_{\text{int}} = 0.015$ eV/Å$^2$ the per-atom interlayer shear stiffness and $n = 19$ the number of interfaces within a 20-layer graphene flake considered in the flexible MD simulations. To suppress unphysical wrinkling during rotation, each flake atom was harmonically restrained in the out-of-plane direction with a spring stiffness of $k_z = 0.167$ eV/Å$^2$. All substrate atoms were constrained to their equilibrium positions by harmonic springs with lateral stiffness of $k_{\text{int}}$ and vertical stiffness of $k_z$.

Quasi-static torsional simulations were performed by incremental rotation of the rigid driving ring followed by full relaxation of the flake and substrate. The angular step ($\Delta\theta$) was chosen such that the maximum displacement of any driving atom satisfied $R \cdot \Delta\theta < 0.1$ Å. Each relaxation used a combination of conjugate-gradient minimization with a relative energy tolerance of $1.0 \times 10^{-20}$ and the FIRE refinement[43] with a force convergence criterion of $1.0 \times 10^{-3}$ eV/Å. In the rigid atomistic calculations, a rigid flake was rotated at an increment of 0.01° while maintained at a fixed interlayer separation of 3.35 Å with the rigid substrate. All MD simulations and atomistic calculations were performed using the LAMMPS package[44].

The resisting torque with respect to the flake center was calculated from the spring forces as

$$T_z = \sum_i [(\boldsymbol{r}_i - \boldsymbol{r}_c) \times \boldsymbol{F}_i] \cdot \boldsymbol{e}_z,$$

where $\boldsymbol{r}_i$ is the atomic position of atom $i$, $\boldsymbol{r}_c$ is the center of mass of the flake, $\boldsymbol{F}_i$ is the spring force acting on atom $i$, and $\boldsymbol{e}_z$ is the unit vector normal to the interface. Per-atom in-plane deviatoric stress ($\sigma$) in the flake was calculated from the virial stress tensor after removing spring contributions,

$$\sigma = \sqrt{\sigma_{xy}^2 + \frac{1}{2}\left(\sigma_{xx} - \sigma_{yy}\right)^2}.$$

**Acknowledgement**

We are grateful to Guitian Qiu, Xuehao Guo, and Yujian Zhang for their assistance in the preparing and fabricating of the stacked vdW mesa samples. D.P. and Q.Z. would like to thank the financial supports by the National Key R&D Program of China (No. 2023YFB4603604), the National Natural Science Foundation of China (Grant No. 12574037), Shenzhen Science and Technology Program (No. KQTD20240729102211015, No. JCYJ20210324100600001), Shenzhen Key Laboratory of Superlubricity Technology (No. ZDSYS20230626091701002), and the National Natural Science Foundation of Guangdong Province (No. 2025A1515012226). X.G. acknowledges the financial supports from the National Natural Science Foundation of China (No. 12402133) and National Key Project (No. MJZ5-2N22). M.U. acknowledges the financial support of the BSF-NSF grant No. 2023614. O.H. is grateful for the generous financial support of the Heinemann Chair in Physical Chemistry and Tel Aviv University Center for Nanoscience and Nanotechnology. The computation work was performed on the supercomputer at the Supercomputing Center of University of Science and Technology of China and Tianhe new generation supercomputer at the National Supercomputer Center in Tianjin.